\documentstyle[12pt]{article}
\newcommand{\lpar}{\stackrel{\leftarrow}{\partial}}
\newcommand{\rpar}{\stackrel{\rightarrow}{\partial}}
\begin{document}
\renewcommand{\thefootnote}{\fnsymbol{footnote}}
\begin{center}
{\large\bf Degenerate Odd Poisson Bracket  \\
on Grassmann Variables \\ }
\vspace{1cm} Vyacheslav A. Soroka
\footnote{E-mail:  vsoroka@kipt.kharkov.ua}
\vspace{1cm}\\
{\it Institute of Theoretical Physics}\\
{\it National Science Center}\\
{\it "Kharkov Institute of Physics and Technology"}\\
{\it 310108, Kharkov, Ukraine}\\
\vspace{1.5cm}
\end{center}
\begin{quotation}
{\small\rm A linear degenerate odd Poisson bracket (antibracket) realized
solely on Grassmann variables is presented. It is revealed that this
bracket has at once three nilpotent $\Delta$-like differential
operators of the first, the second and the third orders with respect to
the Grassmann derivatives. It is shown that these $\Delta$-like
operators together with the Grassmann-odd nilpotent Casimir function
of this bracket form a finite-dimensional Lie superalgebra.}
\end{quotation}

\renewcommand{\thefootnote}{\arabic{footnote}}
\setcounter{footnote}0

\vspace{1cm}
Until now in the main non-degenerate odd Poisson brackets have been
studied and applied both in the Batalin-Vilkovisky formalism
\cite{bv,bv1,blt,bt,sc,kn} for the quantization of gauge theories and in
the description of the Hamiltonian dynamics by means of the odd bracket
\cite{l,vpst,s,k,kn1,vst,vs,s1,s2,n,vty,s3}. However, as it is well-known
on an example of usual even Poisson brackets, that degenerate brackets
play also a very important role in mathematics and in physical
applications (see, for example, \cite{km} and references therein).

In this paper we present a linear degenerate odd Poisson bracket which is
realized solely on the Grassmann variables. We obtained that this bracket,
in contrast with the non-degenerate odd bracket having only one Grassmann-odd
nilpotent differential $\Delta$-operator of the second order, has at once
three Grassmann-odd nilpotent $\Delta$-like differential operators of the
first, the second and the third orders with respect to the Grassmann
derivatives. We show that these $\Delta$-like operators together with the
Grassmann-odd nilpotent Casimir function of this degenerate odd bracket
form a finite-dimensional Lie superalgebra.

There is a well-known linear degenerate even Poisson bracket given
in terms of the commuting (Grassmann-even) variables $X_\alpha$
$$
\{X_\alpha, X_\beta \}_0 = \varepsilon_{\alpha\beta\gamma}X_\gamma;\qquad
(\alpha,\beta,\gamma = 1,2,3),
\eqno {(1)}
$$
where $\varepsilon_{\alpha\beta\gamma}$ is the Levi-Civita tensor.
The linear even Poisson brackets like (1) play a very important role in the
theory of Lie algebras, Lie groups and their representations (see, for
example, \cite{km,b}). In general the degenerate Poisson bracket has
Casimir functions $C_k(X)$ whose brackets with any function $f(X)$ vanish
$$
\{C_k(X), f(X)\}_0 = 0.
$$
As a rule level surfaces $C_k(X) = const.$ all of the independent Casimir
functions define symplectic leaves on which the bracket becomes
non-degenerate and a closed 2-form (symplectic form) can be defined. The
bracket (1) has only one Casimir function
$$
C = \sum^{3}_{\alpha=1} (X_\alpha)^2
$$
and the symplectic leaves in the form of the $S^2$-spheres of a definite
radius.

Now let us replace in (1) the commuting variables $X_\alpha$ by Grassmann
variables $\theta_\alpha$. Then we obtain a binary composition
$$
\{\theta_\alpha, \theta_\beta \}_1 =
\varepsilon_{\alpha\beta\gamma}\theta_\gamma;\qquad (\alpha,\beta,\gamma =
1,2,3),
\eqno {(2)}
$$
which meets all the properties of the odd Poisson brackets:
$$
\{ A, B + C \}_1 = \{ A, B \}_1 + \{ A, C \}_1\ ,
\eqno {(3)}
$$
$$
g(\{ A, B \}_1) = g(A) + g(B) + 1 \pmod 2\ ,
\eqno {(4)}
$$
$$
\{ A , B \}_1 = -(-1) ^{(g(A) + 1)(g(B) + 1)} \{ B , A \}_1\ ,
\eqno {(5)}
$$
$$
\sum_{(ABC)}(-1) ^{(g(A) + 1)(g(C) + 1)} \{ A , \{ B , C \}_1 \}_1 = 0\ ,
\eqno {(6)}
$$
$$
\{ A , B C \}_1 = \{ A , B \}_1\ C +
(- 1) ^{(g(A) + 1)g(B) }\ B \{ A , C \}_1\ ,
\eqno {(7)}
$$
where $A, B, C$ are functions of phase variables $\theta_\alpha$, $g(A)$
is a Grassmann parity of the quantity $A$ and a sum with the symbol
$(ABC)$ in (6) denotes a summation over cyclic permutations of the
quantities $A, B, C$. In order to establish the validity of the Jacobi
identities (6) for the bracket (2) we need to use the following relation
for the completely antisymmetric tensor $\varepsilon_{\alpha\beta\gamma}$
$$
\varepsilon_{\alpha\beta\lambda}\varepsilon_{\gamma\delta\lambda} =
\delta_{\alpha\gamma}\delta_{\beta\delta} -
\delta_{\alpha\delta}\delta_{\beta\gamma}.
\eqno {(8)}
$$
It is surprising enough that the odd bracket can be realized solely on
the Grassmann variables as well as an even Martin bracket.

Using relation (8) we can verify that the degenerate odd bracket (2)
has the following Grassmann-odd nilpotent Casimir function
$$
\Delta_{+3} = {1\over\sqrt{3!}} \varepsilon_{\alpha\beta\gamma}
\theta_\alpha \theta_\beta\theta_\gamma\ ,\qquad
\{\Delta_{+3}, ...\}_1 = 0\ ,\qquad(\Delta_{+3})^2 = 0\ .
\eqno {(9)}
$$
Such a notation for this Casimir function will be clear below.

By the way, let us note that with the use of the completely
antisymmetric five-tensor $\varepsilon_{\alpha\beta\gamma\delta\lambda}$
$(\alpha,...,\lambda = 1,...,5)$ we can also build only in terms of
Grassmann variables a non-linear degenerate odd Poisson bracket of the
following form
$$
\{\theta_\alpha, \theta_\beta \}_1 =
\varepsilon_{\alpha\beta\gamma\delta\lambda}
\theta_\gamma\theta_\delta\theta_\lambda\ .
\eqno {(10)}
$$
Indeed, with the use of the relations for the five-tensor
$\varepsilon_{\alpha\beta\gamma\delta\lambda}$, which are similar to (8),
we can establish for the bracket (10) the relation
$$
\{\theta_\alpha, \{\theta_\beta, \theta_\gamma \}_1 \}_1 = 0
$$
and therefore the validity of the Jacobi identities (6). The rest of
the odd bracket properties (3)-(5) and (7) are evidently fulfilled
for the bracket (10). The odd bracket (10) has several nilpotent
Casimir functions:
$$
C = \varepsilon_{\alpha\beta\gamma\delta\lambda}
\theta_\alpha\theta_\beta\theta_\gamma\theta_\delta\theta_\lambda\ ,\qquad
\{C, ...\}_1 = 0\ ,\qquad C^2 = 0\ ;
$$
$$
C_\alpha = \varepsilon_{\alpha\beta\gamma\delta\lambda}
\theta_\beta\theta_\gamma\theta_\delta\theta_\lambda\ ,\qquad
\{C_\alpha, ...\}_1 = 0\ ,\qquad (C_\alpha)^2 = 0\ ;
$$
$$
C_{\alpha\beta} = \varepsilon_{\alpha\beta\gamma\delta\lambda}
\theta_\gamma\theta_\delta\theta_\lambda\ ,\qquad
\{C_{\alpha\beta}, ...\}_1 = 0\ ,\qquad (C_{\alpha\beta})^2 = 0\ ,
$$
where no summations in the indices are assumed in the nilpotency
conditions.

It is a well-known fact that, in contrast to the even Poisson bracket, in
the case of the odd Poisson bracket it can be built a Grassmann-odd
nilpotent differential $\Delta$-operator of the second order, which has
naturally appeared in the Batalin-Vilkovisky scheme
\cite{bv,bv1,blt,bt,sc,kn} for the quantization of gauge theories in
the Lagrangian approach. This operator plays also a very important role
in the formulation of the Hamiltonian dynamics by means of the odd Poisson
bracket with the help of the Grassmann-odd Hamiltonian
$\bar H$ $(g(\bar H) = 1)$ \cite{l,vpst,s,k,kn1,s2}
$$
{dx^A\over dt} = \{ x^A, \bar H \}_{1}\ ,
\eqno{(11)}
$$
where $t$ is the time and $x^A = (x^i, \theta_i)$  $( i = 1,...,n )$ are
the canonical phase coordinates. In the Hamiltonian dynamics expressed in
terms of the odd Poisson bracket the $\Delta$-operator can be used to
distinguish the non-dissipative dynamical systems from the dissipative
ones. In fact, for the usual non-degenerate odd Poisson bracket in the
canonical form
$$
\{ A , B \}_1 = A \sum_{i = 1}^n
\left(\lpar_{x^i}
\rpar_{\theta_i} -
\lpar_{\theta_i}
\rpar_{x^i}\right) B\ ,
$$
where $\lpar$ and $\rpar$ are the right and left derivatives and
the notation $\partial_{x^A} \equiv {\partial \over {\partial x^A}}$ is
introduced,
$$
\Delta = 2 \sum_{i = 1}^n \partial_{x^i} \partial_{\theta_i}
$$
and the infinitesimal form of the Liouville theorem is
$$
sTr (\partial_{x^A} \{ x^B, \bar H \}_{1}) \equiv  (-1)^{g(x^A)}
\partial_{x^A} (\{ x^A, \bar H \}_{1}) = \Delta \bar H = 0\ .
$$
If $\Delta \bar H \neq 0$, then the Liouville theorem takes no place, and
a dynamical system described by means of the Hamilton equation in terms
of the odd Poisson bracket (11) with such a Hamiltonian is dissipative.

Now let us try to build the $\Delta$-operator for the linear degenerate
odd bracket (2). It is amusing that we are able to build at once three
$\Delta$-like Grassmann-odd nilpotent operators which are the differential
operators of the first, the second and the third orders respectively
$$
\Delta_{+1} = {1\over\sqrt{2}} \theta_\alpha \theta_\beta
\varepsilon_{\alpha\beta\gamma} \partial_{\theta_\gamma}
\ ,\qquad (\Delta_{+1})^2 = 0\ ;
\eqno {(12)}
$$
$$
\Delta_{-1} = {1\over\sqrt{2}} \theta_\alpha
\varepsilon_{\alpha\beta\gamma}
\partial_{\theta_\beta} \partial_{\theta_\gamma}
\ ,\qquad (\Delta_{-1})^2 = 0\ ;
\eqno {(13)}
$$
$$
\Delta_{-3} = {1\over\sqrt{3!}} \varepsilon_{\alpha\beta\gamma}
\partial_{\theta_\alpha} \partial_{\theta_\beta} \partial_{\theta_\gamma}
\ ,\qquad (\Delta_{-3})^2 = 0\ .
\eqno {(14)}
$$
It is also a surprise to reveal that these $\Delta$-like operators
together with the Casimir function $\Delta_{+3}$ (9) are closed into the
finite-dimensional Lie superalgebra in which the anticommuting relations
between the quantities $\Delta_\lambda$ $(\lambda = -3, -1, +1, +3 )$ with
the nonzero right-hand side are\footnote{In order to avoid the confusion
let us note that below $[A, B] = AB -BA$ and $\{ A, B\} = AB + BA$.}
$$
\{ \Delta_{-1}, \Delta_{+1} \} = Z\ ;
\eqno {(15)}
$$
$$
\{ \Delta_{-3}, \Delta_{+3} \} = -6 - 3Z\ ,
\eqno {(16)}
$$
where $$ Z = D^2 - 3D
\eqno {(17)}
$$
is a center of this superalgebra
$$
[ Z, \Delta_\lambda ] = 0\ ,\qquad (\lambda = -3, -1, +1, +3)
\eqno {(18)}
$$
and
$$
D = \theta_\alpha \partial_{\theta_\alpha}
\eqno {(19)}
$$
is a "dilatation" operator for the Grassmann variables
$\theta_\alpha$ which distinguishes the $\Delta_\lambda$-operators
with respect to their uniformity degrees in $\theta$
$$
[ D, \Delta_\lambda ] = \lambda
\Delta_\lambda\ ,\qquad (\lambda = -3, -1, +1, +3)\ .
\eqno {(20)}
$$
We can add to this superalgebra the generators $S_\alpha$ of rotations in
the $\theta$-space
$$
S_\alpha = \theta_\gamma \varepsilon_{\alpha\beta\gamma}
\partial_{\theta_\beta}
\eqno {(21)}
$$
with the following commutation relations:
$$
[ S_\alpha, S_\beta ] = \varepsilon_{\alpha\beta\gamma} S_\gamma\ ,\qquad
[ S_\alpha, \Delta_\lambda ] = 0\ ,
\eqno {(22a,b)}
$$
$$
[ S_\alpha, Z ] = 0\ ,\qquad [ S_\alpha, D ] = 0\ .
\eqno {(22c,d)}
$$
In order to prove the nilpotency of the operators $\Delta_{+1}$ and
$\Delta_{-1}$ and to establish most of the permutation relations
for the Lie superalgebra (15)-(22) we have to use the relations (8) for
the Levi-Civita tensor $\varepsilon_{\alpha\beta\gamma}$.
Note that the center $Z$ (17) coincides with the expression for the
quadratic Casimir operator of the Lie algebra (22a) for the generators
$S_\alpha$ given in the representation (21)
$$
S_\alpha S_\alpha = Z\ .
\eqno {(23)}
$$

\bigskip
The author is thankful to V.D. Gershun for useful discussions and is
sincerely grateful to A.A. Kashinsky, V.E. Korol', D.V. Soroka and
A.A. Zheltukhin for their warm care during his illness.

\medskip
This work was supported, in part, by the Ukrainian State Foundation
of Fundamental Researches, Grant No 2.5.1/54, by Grant INTAS No 93-127
(Extension) and by Grant INTAS No 93-633 (Extension).

\end{document}